\documentclass{emulateapj}
\def\stacksymbols #1#2#3#4{\def\theguybelow{#2}
        \def\verticalposition{\lower#3pt}
        \def\spacingwithinsymbol{\baselineskip0pt\lineskip#4pt}
        \mathrel{\mathpalette\intermediary#1}}
\def\intermediary #1#2{\verticalposition\vbox{\spacingwithinsymbol
        \everycr={}\tabskip0pt
        \halign{$\mathsurround0pt#1\hfil##\hfil$\crcr#2\crcr
                \theguybelow\crcr}}}
\def\lta{\stacksymbols{<}{\sim}{2.5}{.2}}
\def\gta{\stacksymbols{>}{\sim}{3}{.5}}

\begin{document}

\title{LARGE SCATTER IN X-RAY EMISSION AMONG ELLIPTICAL GALAXIES:
CORRELATIONS WITH MASS OF HOST HALO}

\author{William G. Mathews\footnotemark[1],
Fabrizio Brighenti{\footnotemark[1]$^,$\footnotemark[2]},
Andreas Faltenbacher\footnotemark[1],
David A. Buote\footnotemark[3],
Philip J. Humphrey\footnotemark[3],
Fabio Gastaldello\footnotemark[3],
\& Luca Zappacosta\footnotemark[3]}

\footnotetext[1]{UCO/Lick Observatory,
Dept. of Astronomy and Astrophysics,
University of California, Santa Cruz, CA 95064}

\footnotetext[2]{Dipartimento di Astronomia,
Universit\`a di Bologna,
via Ranzani 1,
Bologna 40127, Italy}

\footnotetext[3]{Dept. of Physics \& Astronomy, Univ. California 
Irvine, 4129 Fredreick Reines Hall, Irvine, CA 92697}

\begin{abstract}
Optically similar elliptical galaxies have an enormous range
of X-ray luminosities.
We show that 
this range can be attributed to large variations 
in the dark halo mass $M_{vir}$ determined from X-ray observations.
The K-band luminosity of ellipticals varies  
with virial mass, 
$L_K \propto M_{vir}^{0.75 \pm 0.22}$, 
but with considerable scatter, probably due to the stochastic 
incidence of massive satellite galaxies that merge 
by dynamical friction to form group-centered ellipticals. 
Both the observed X-ray luminosity $L_x \propto M_{vir}^{2.4}$ and 
$L_x/L_K \propto M_{vir}^{1.6}$ are sufficiently sensitive 
to the virial mass 
to explain the wide variation observed in $L_x$ among 
galaxies of similar $L_K$.
The central galaxy supernova energy per particle of diffuse gas 
increases dramatically with decreasing virial mass and 
elliptical galaxies with the lowest X-ray luminosities 
(and $M_{vir}$) are easily 
explained by supernova-driven outflows. 
\end{abstract}


\section*{}
\section{INTRODUCTION AND BACKGROUND}

In spite of their very regular optical properties, 
elliptical galaxies with similar optical luminosities have 
X-ray luminosities that range over more than two orders of magnitude.
Many attempts to explain this huge scatter or to find observable 
properties that correlate with residuals in this scatter 
have been inconclusive 
(e.g. Eskridge, Fabbiano \& Kim 1995a,b,c;
Ellis \& O'Sullivan 2006;
Mathews \& Brighenti 2003 for a review).

Figure 1 shows a plot of X-ray and K-band 
luminosities for elliptical galaxies taken from the 
catalog of Ellis \& O'Sullivan (2006). 
Most of the X-ray emission is from diffuse hot gas 
although stellar emission from low mass X-ray binaries (LMXB) 
contributes in galaxies of the lowest $L_x$. 
While the lower envelope of the distribution is well-defined 
by the locus for stellar LMXB emission, 
X-ray images of these low-$L_x$ elliptical galaxies also show 
definite evidence of diffuse gas emission 
(e.g. Sarazin et al. 2001). 
There is no apparent concentration of gas-free ellipticals 
along the LMXB locus, confirming the general prevalence of diffuse
emission. 
Figure 1 includes only a few ram-stripped, low-$L_x$ elliptical 
galaxies orbiting in massive clusters that have nevertheless retained 
a small mass of $\sim 1$keV gas 
(e.g. Sun et al. 2004; 2005). 
Our interest here is not these strongly ram-stripped galaxies, 
but similar galaxies that are isolated 
or central galaxies in groups.

Using an early data set based on {\it Einstein}, ROSAT 
and ASCA observations, 
Mathews \& Brighenti (1998a) showed that the radius 
of half-X-ray emission $r_{effx}$ decreases systematically 
with decreasing $L_x$ in Figure 1. 
It was demonstrated later that this correlation could 
be understood as a variation of the dark halo mass 
(Mathews \& Brighenti 1998b). 
In this paper we extend this idea using more recent data 
to show that the scatter in $L_x$ is intimately related 
to stochastic and systematic variations of the virial mass 
$M_{vir}$ of the dark 
halos that surround them. 
Galaxies with smaller $M_{vir}$ necessarily have fewer 
baryons and lower $L_x$ and we use simple galaxy plus 
halo models to show that this effect 
alone can explain at most about half of the total observed 
range in $\log L_x$. 
However, galaxies/groups with the lowest $L_x$ require 
AGN or supernova-driven outflows.

\section{CORRELATIONS AMONG $L_x$, $L_K$ AND $M_{vir}$}

The virial mass of dark host halos around 
elliptical galaxies can be directly determined 
from radial temperature and density profiles in 
the hot gas, assuming hydrostatic equilibrium.
The total mass based on {\it Chandra} and XMM images
can be attributed to a combination of stellar and 
dark mass profiles.
{\it Chandra} observations are particularly 
valuable since LMXB sources can be resolved and removed.
The galaxy sample we consider here with known 
viral masses is shown in Figure 1 
with filled symbols: squares are from 
Humphrey et al (2006), circles are from 
Gastaldello et al. (2006) and the hexagons 
are X-ray luminous galaxies tabulated in Mathews et al. (2005).

Figure 2a shows that $M_{vir}$ for the sample galaxies 
increases with $L_x$ according to 
the correlation 
\begin{equation}
\log M_{vir} = -(4.20 \pm 0.92) + (0.42 \pm 0.02)\log L_x
\end{equation}
(with correlation coefficient 0.75)
indicated with a dashed line, 
i.e. $L_x \propto M_{vir}^{2.4}$.
This and subsequent correlations are formal least square 
fits, assuming a mean typical error 
$\Delta \log M_{vir} = 0.090$ which is small 
compared to the intrinsic spread in the data. 
The quality of our least squares fit in Figure 2a 
is limited by the cosmic 
spread in $M_{vir}$, but it is consistent with 
a slightly steeper variation 
$ M_{vir} \propto L_x^{0.65}$ in massive clusters 
found by Reiprich \& B\"ohringer (2002).

Figure 2b indicates a very wide variation in 
the virial mass with K-band luminosity of the central 
galaxy $L_K$, superimposed on a 
positive $M_{vir}$-$L_K$ correlation,
{\begin{equation}
\log M_{vir} = -(1.75 \pm 1.04) + (1.34 \pm 0.09) \log L_K, 
\end{equation}
(dashed line in Fig 2b) with correlation coefficient 0.58.

The mean optical luminosity of group and cluster-centered
elliptical galaxies is known to increase, then
saturate as the virial mass of the cluster increases.
Halo masses around elliptical galaxies
in small groups can be determined
from the SDSS galaxy-mass correlation function
(Cooray and Milosavljevic 2005),
from weak galaxy-galaxy lensing 
(Yang et al. 2003), and from the X-ray temperature of the gas
(Lin \& Mohr 2004). 
The $L_K$ of the central galaxy varies as $M_{vir}^{0.85}$ for 
$M_{vir} \lta 10^{12}$ $M_{\odot}$ and as $M_{vir}^{0.25}$
for $M_{vir} \gta 10^{14.5}$ $M_{\odot}$.
The intermediate variation $L_K \propto M_{vir}^{0.75}$ 
in equation (2) is consistent with these results 
for the range in $M_{vir}$ that we consider.

The scatter in 
Figure 2b arises because of
an intrinsic (log-normal) cosmic variance of $M_{vir}$ 
among galaxies of similar $L_K$.
The central galaxy
forms from a rather small number of luminous
satellite galaxies that merge at the halo center 
because of dynamical friction 
(e.g. Faltenbacher \& Mathews 2005).
Since dynamical friction varies as the square of the 
ratio of the galaxy subhalo 
mass to the mass of the host halo, only the most luminous 
group satellite galaxies merge at the center. 
Therefore, because of the relative rarity and stochastic 
incidence of luminous satellite galaxies, 
halos of the same mass can have a range of $L_K$ for 
the central elliptical 
(e.g. Cooray 2006). 
Conversely, 
halos of different $M_{vir}$ can contain 
central galaxies with similar $L_K$,
as seen in Figure 2b. 

Figure 2c shows that the mean gas temperature 
$\langle k T \rangle$ 
(estimated from $T(r)$ profiles for the Humphrey et al 
2006 data) 
is tightly correlated with the 
virial mass. 
The dashed line describing this correlation 
is given by 
\begin{equation}
\log M_{vir} = (13.5 \pm 0.02) + (2.07 \pm 0.09) 
\log \langle kT \rangle.
\end{equation}
This $kT$-$M_{vir}$ correlation is similar to those 
found in clusters with more massive halos 
(Arnaud et al. 2005; Vikhlinin et al. 2006).

\section{RELATION OF $L_x$ TO BARYON CLOSURE}

The physical origin of the 
vertical deviation of $L_x$ in Figure 1 is best characterized 
by the logarithmic distance $D(\log L_x)$ of each galaxy from 
the (dash-dotted) locus of baryonic closure, 
the approximate $L_x$ expected if the 
most massive dark 
halos for each $L_K$ are filled with gas (and stars) containing the 
maximum baryonic mass allowed by the cosmic baryon ratio
$f_b = \Omega_b/\Omega_m = 0.176$ observed by 
the {\it Wilkinson Microwave Anisotropy Probe} 
(WMAP; Spergel et al. 2006).
If a galaxy/group is baryonically closed, 
no gas has been ejected from the confining halo by AGN or 
supernova energy. 
Since the total mass of diffuse gas can be reasonably 
well estimated in 
several of the most luminous groups in the sample, 
we know that at least some 
galaxies/groups with the largest $L_x$ are quite close to 
the closure locus (Mathews et al. 2005),
\begin{displaymath}
(\log L_x)_{closure} \approx 18.361 + 2.18 \log L_K, 
\end{displaymath} 
shown by the dash-dotted line in Figure 1.
The deviation in $L_x$ of a given galaxy below the closure
locus is
\begin{equation}
D(\log L_x) = (\log L_x)_{closure} - \log L_x.
\end{equation}

Figure 2d shows
an anticorrelation between $D(\log L_x)$ and $M_{vir}$
\begin{equation}
\log M_{vir} = (14.01 \pm 0.03) - (0.34 \pm 0.03) D(\log L_x),
\end{equation}
(dashed line in Fig. 2d),
confirming that the virial mass is key to understanding the
wide scatter in $L_x$ among elliptical galaxies in Figure 1. 
Eliminating $M_{vir}$ between equations (1) and (2), 
we derive the approximate direction in the $L_x$-$L_K$ plane 
along which $M_{vir}$ decreases most rapidly, 
shown by the short-dashed arrow in Fig. 1.

\section{ISOTHERMAL MODELS}

The closure locus in Figure 1 -- the approximate 
maximum $M_{vir}$ and $L_x$ for galaxy/groups for each $L_K$ -- 
can be related to simple models for the hot gas. 
We assume that the gas is isothermal with $\langle kT \rangle$ 
given by equation (3) and in hydrostatic equilibrium in the 
potentials of the dark halo and central galaxy. 
The NFW halo is characterized by its virial 
mass $M_{vir}$ (using a mean concentration $c(M_{vir})$ 
from Bullock et al. 2001). 
The virial radius is defined by an overdensity $\Delta = 104$ 
above the 
critical density $\rho_c = 9.24\times 10^{-30}$ gm cm$^{-3}$.
The central 
galaxy is assumed to have a de Vaucouleurs mass distribution 
described by $L_K$ and effective radius 
$r_{eff} = 4.16 (M_*/10^{11}~M_{\odot})^{0.56}$ kpc 
(Boylan-Kolchin et al. 2005).
The B-band stellar mass to light ratio is 
$\Upsilon_B = M_*/L_B = 7.5$ and we adopt 
$L_K/L_{K\odot} = 4.15 L_B/L_{B\odot}$ 
appropriate for an old stellar population 
in elliptical galaxies (Ellis \& O'Sullivan 2006). 
Solutions of these isothermal, hydrostatic models with 
$f_b = 0.176$ (and including gas self-gravity) 
give the maximum 
$L_x$ within $r_{vir}$ for any given $M_{vir}$ and $L_K$.
For each $L_K$ we increase $M_{vir}$ until $L_x$
of the model exceeds all observed values and eventually 
recover the approximate closure locus 
described by Mathews et al. (2005) 
(dash-dotted line in Figure 1), given by 
$L_x = 2.19 \times 10^{42} (L_K/10^{11}~L_{K\odot})^{2.18}$
erg s$^{-1}$. 
The closure locus empirically defines an upper limit to 
the $L_x$-$L_K$ distribution and is consistent with the 
large baryon fractions $\approx f_b$ determined for 
well-observed galaxy/groups having the largest $L_x/L_K$.
Since $L_x$ of the models extend to $r_{vir}$ while the 
observed $L_x$ do not, 
even galaxies with the largest $L_x$ in Figure 1 
are expected to lie slightly below the closure locus.

In principle, galaxies that lie significantly 
below the closure line in Figure 1 could 
also be baryonically closed, i.e. with fixed 
$f_b = 0.176$, $L_x$ is expected to decrease with $M_{vir}$. 
Although we view this possibility as highly unlikely, 
it is interesting to explore how much of the spread in 
Figure 1 can be understood with this extreme assumption 
of universal baryon closure.
Consider simple models in which $M_{vir}$ decreases 
below the closure locus in Figure 1, 
holding $L_K$ of the 
central galaxy fixed and ignoring the stellar mass of 
satellite galaxies.
For central galaxies of mass 
$M_* = \Upsilon_B L_B = (7.5/4.15) L_K$ $M_{\odot}$ 
the fraction of baryons remaining in diffuse gas is 
\begin{equation}
f_{g} = M_{gas}/M_{vir} = f_b - 1.8{ L_K / M_{vir}}.
\end{equation}
$L_x$ declines with $M_{vir}$ faster than $M_{gas}$ since
stars (with constant $L_K$) 
consume an increasingly larger fraction of the baryons
in smaller halos.
In this simple model, where mass loss from evolving stars 
is neglected, the mass of diffuse gas (and $L_x$) 
completely vanishes ($f_g = 0$) for halo masses less than 
$M_{vir,0} = 10.3L_K$.

Isothermal models with $\log L_K = 11.5$ 
and $\log (M_{vir}/M_{\odot}) = 13.5$, 13 and 12.6 
(models $a$, $b$, and $c$ respectively) are listed in Table 1
and plotted in Figure 1. 
From equation (6) 
model $c$ is close to the $f_g = 0$ limit. 
The X-ray luminosities of these models extend over 
nearly half the total spread in Figure 1.
The radius $r_{effx}$ in Table 1 
that includes half of the total projected
$L_x$ decreases systematically with 
$L_x$, qualitatively confirming the correlation
between the observed X-ray sizes and
$L_x$ found by Mathews \& Brighenti (1998a).
Also listed in Table 1 are the thermal and binding energies 
per gas particle, $e_{th}$ and $e_{bind}$, as well as the 
corresponding energy released by supernovae of Type II,
$e_{II}$, calculated assuming a Salpeter IMF, and Type Ia, 
$e_{Ia}$, estimated from the total iron mass produced by 
an old stellar population, $M_{Fe}/M_{*} \approx 0.0006$ 
(De Grandi et al. 2004), assuming 0.7 $M_{\odot}$ in iron per 
supernova.

As $M_{vir}$ decreases, the supernova energy 
in the central galaxy varies from 
relative insignificance for model $a$ to dominance in 
model $c$ where $e_{II} + e_{Ia} \gg e_{bind}$,
the gravitational binding energy per gas particle.
The global energetics in Table 1 strongly suggest that 
the remaining scatter in $L_x$ in Figure 
1 below model $c$ 
can easily be explained by supernova-driven gaseous outflows from 
low $M_{vir}$ halos.  
It is more likely that outflows reduce 
$L_x$ for all galaxies in Figure 1, 
and that outflows are disproportionally effective as 
$M_{vir}$ decreases with increasing $D(\log L_x)$ 
(Fig. 2d). 

Another feature in the galaxy distribution in Figure 1 
that remains to be understood is the absence of 
galaxies near the closure locus but with 
$\log L_K \lta 11.2$. 
If the halo mass decreases in the direction of the 
short-dashed arrow, 
it is plausible that the absence of galaxies in this 
region can also be understood in terms of 
supernova-driven outflows 
combined with a modest gas supply from evolving stars, 
maintaining $L_x$ close to the LMXB lower limit. 

Finally, we note another observable feature that 
correlates with $M_{vir}$: 
the shape of the 
radial gas temperature profile $T(r)$. 
Massive galaxy groups with large $L_x$ 
have a temperature maximum displaced from the center, 
but in low-$L_x$ ellipticals the temperature peaks at the center 
(e.g. Humphrey et al. 2004; 
Humphrey et al. 2006; O'Sullivan \& Ponman 2004). 
The origin of this difference in thermal profiles can 
be explained by the ratio of the mass of the galaxy 
to its host halo (Mathews \& Brighenti 2003).
When a group-centered elliptical resides in a more massive
halo, the virial (and gas) temperature throughout most of the halo 
exceeds that of the central galaxy, resulting in a 
positive temperature gradient in the inner regions.
However, when a similar elliptical has a halo of lower mass, 
the gas temperature peaks at the center where it 
approaches the virial temperature of the stellar bulge. 
This observed pattern is consistent
with the expected variation of $L_x$ with
$M_{vir}$ in Figures 1 and 2a.

\section{CONCLUSIONS}

Our main conclusions are:
\vskip.1in
\noindent
(1)  The large variation of X-ray luminosity 
among group-centered elliptical galaxies 
of similar optical luminosity can be ascribed to 
a large cosmic variation in the surrounding dark halo mass.

\noindent
(2) For galaxies with similar $L_K$ 
a significant fraction of the observed
correlation between $L_x$ and $M_{vir}$ can be expected 
simply because smaller halos contain fewer baryons, less hot gas, 
and lower $L_x$. 
However, with decreasing $M_{vir}/L_K$ 
supernova (or AGN) driven outflows become more likely, reducing 
$L_x$ further. 

\noindent
(3) $M_{vir}$ decreases with both $L_x$ and $L_K$ in the 
$L_x$-$L_K$ plot.

\noindent
(4) The upper envelope of elliptical galaxies in the 
$L_x$-$L_K$ plot lies close to the locus of WMAP baryon 
closure for galaxies having the most massive dark halos 
at each $L_K$.

\noindent
(5) Previously found correlations between the X-ray luminosity of 
elliptical galaxies (having similar $L_K$) and their X-ray sizes 
(Mathews \& Brighenti 1998a) or temperature profiles 
(Humphrey et al. 2006) are also 
consistent with a systematic decrease in halo mass for 
smaller $L_x$. 




\vskip.1in

Studies of the evolution of hot gas in elliptical galaxies
at UC Santa Cruz are supported by
NASA grants NAG 5-8409 \& ATP02-0122-0079 and NSF grant
AST-0098351 for which we are very grateful.


\clearpage

\begin{deluxetable}{lccc}
\tabletypesize{\scriptsize}
\tablecolumns{4}
\tablewidth{0pc}
\tablecaption{ISOTHERMAL MODELS WITH $L_K = 10^{11.5}$}
\tablehead{
\colhead{} &
\colhead{} &
\colhead{} \\
}
\startdata
model & $a$ & $b$ & $c$ \\
$\log M_{vir} (M_{\odot})$ & 13.5 & 13.0 & 12.6 \\
$\log kT$ (keV) & 1.0 & 0.6 & 0.4 \\
$\log M_{gas} (M_{\odot})$ & 12.66 & 12.07 & 11.11 \\
$\log L_x$ (erg/s) & 43.43 & 42.77 & 42.09 \\
$r_{effx}$ (kpc) & 151 & 104 & 0.9 \\
$e_{th}$ (keV/part) & 1.5 & 0.9 & 0.6 \\
$e_{bind}$ (keV/part) & 2.3 & 1.1 & 0.6 \\
$e_{II}$ (keV/part) & 0.4 & 1.5 & 14. \\
$e_{Ia}$ (keV/part) & 0.03 & 0.1 & 1.2 \\
\enddata
\end{deluxetable}


\clearpage
\begin{figure}
\centering
\includegraphics[bb=90 166 522 519,scale=0.8,angle= 270]
{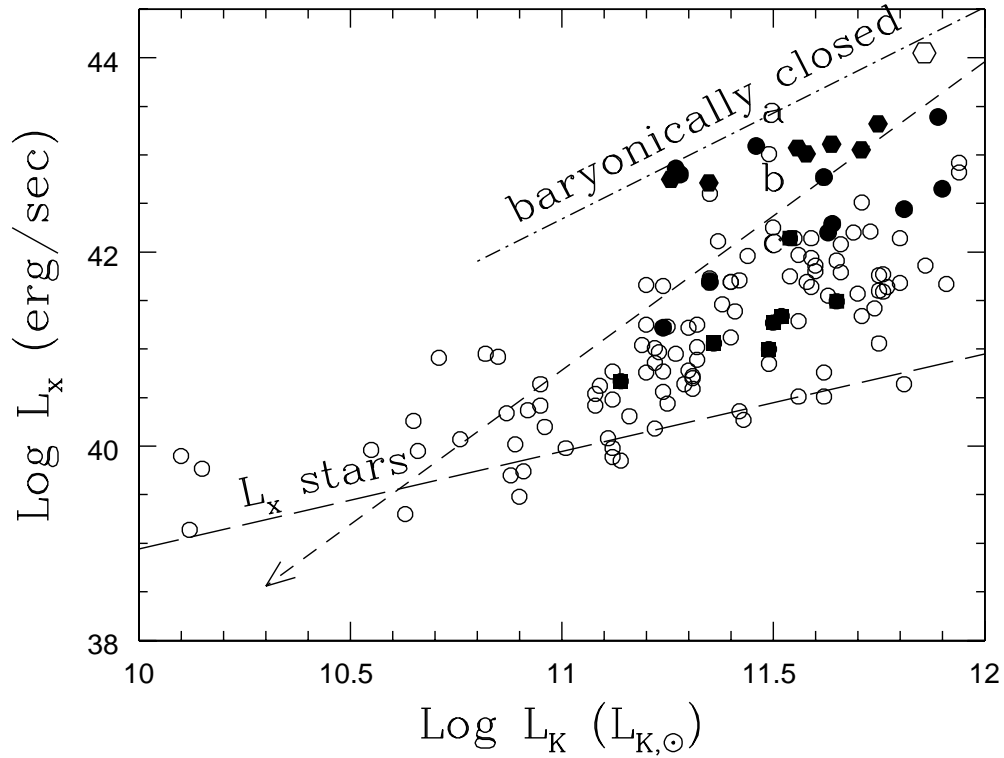}
\vskip.7in
\caption{
K-band and X-ray luminosities of X-ray-detected 
elliptical galaxies ($T < -4$) from the catalog of 
Ellis \& O'Sullivan (2006). 
Filled squares, circles and hexagons are our sample 
galaxies for which the virial mass has been determined 
except RXJ146 
(the open hexagon) which 
is useful to establish the upper envelope of the data.
The lower long-dashed line shows the 
approximate emission from low mass X-ray binary stars 
(Kim \& Fabbiano 2004). The upper dash-dotted line is the 
approximate locus of baryon closure 
(Mathews et al. 2005). 
The short-dashed arrow shows the direction in which $M_{vir}$ 
decreases most rapidly.
Labels $a$, $b$ and $c$ show locations of 
several baryonically closed, isothermal models 
listed in Table 1.
}
\label{f1}
\end{figure}

\clearpage
\begin{figure}
\centering
\vskip1.in
\includegraphics[bb=90 266 522 619,scale=0.8,angle= 270]
{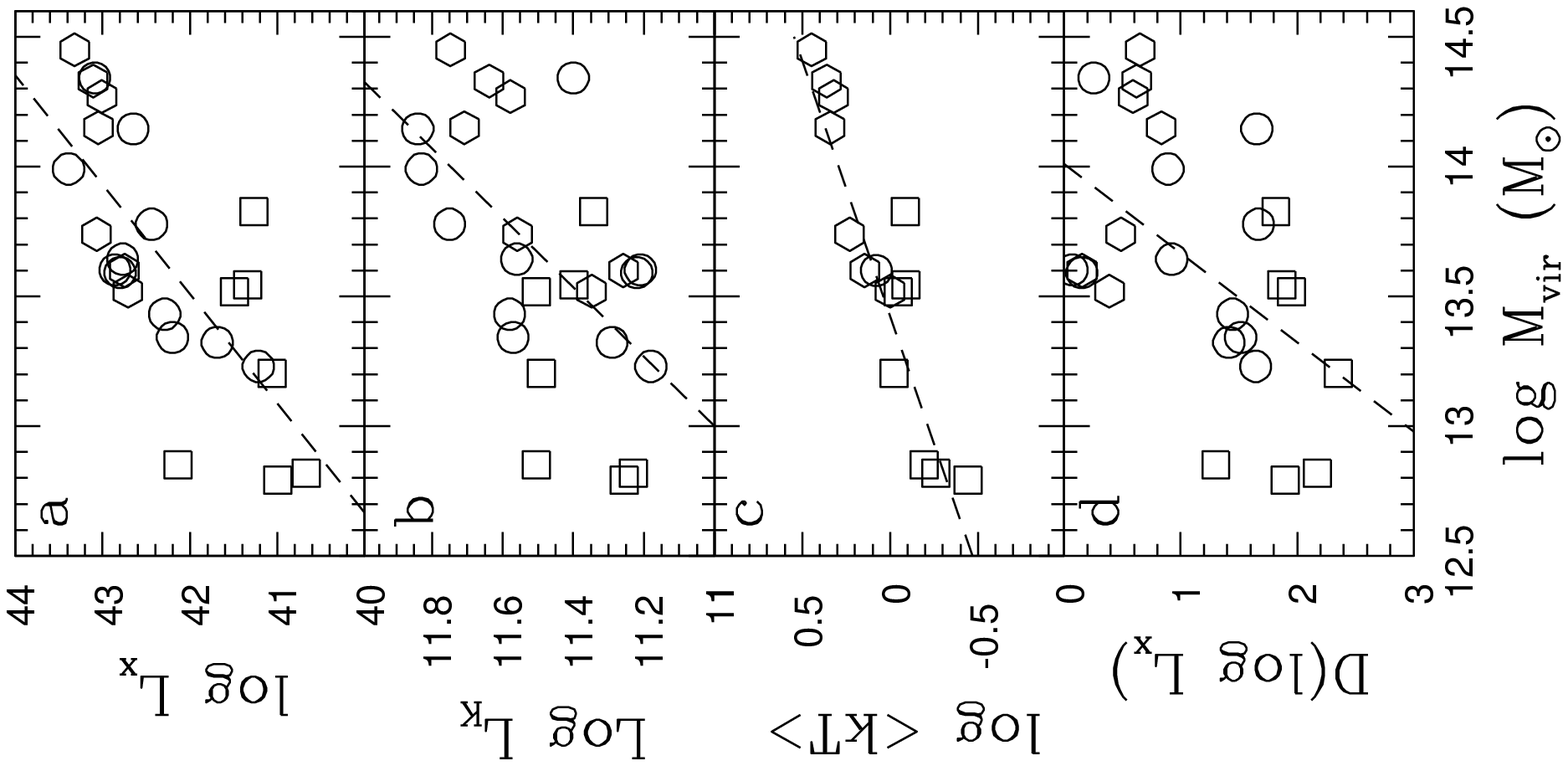}
\vskip.7in
\caption{
(a) Variation of X-ray luminosity $L_x$ (in erg s$^{-1}$)
of the sample galaxies with the virial mass (in $M_{\odot}$)
of their host halos.
(b) Variation of the central galaxy K-band luminosity
(in $L_{K\odot}$) with virial mass.
(c) Variation of mean gas temperature (in keV)
with virial mass.
(d) Variation of the logarithmic deviation from
the locus of baryon closure in Figure 1 with the halo
virial mass. Note that $D(\log L_x)$ increases downward,
as in Figure 1.
Dashed lines show correlations discussed in the text.
Note that all correlations are apparent in published
data from Humphrey et al (2006) ({\it open squares})
and Mathews et al. (2005) ({\it open hexagons}),
while the data of
Gastaldello et al. (2006) ({\it open circles}) fall in 
between.
}
\label{f2}
\end{figure}


\begin{references}


\reference{} Arnaud, M., Pointecouteau, E., \& Pratt, G. W. 2006,
A\&A, 441, 893

\reference{} Boylan-Kolchin, M., Ma, C.-P., \& Quataert, E. 2004, ApJ, 613, L37



\reference{} Bullock, J. et al. 2001, MNRAS, 321, 559

\reference{} Cappellaro, E., Evans, R., \& 
Truratto, M. 1999, A\&A, 350, 459

\reference{} Cooray, A. 2006, MNRAS (submitted) (astro-ph/0509033)

\reference{} Cooray, A. \& Milosavljevic, M. 2005, ApJ, 627, L85


\reference{} De Grandi, S., Ettori, S. \& Longhetti, M. 2004, A\&A, 419, 7

\reference{} Ellis, S. C., \& O'Sullivan, E. 2006, MNRAS, 367, 627

Eskridge, P. B., Fabbiano, G. \& Kim, D.-W. 1995a, ApJS, 97, 141

Eskridge, P. B., Fabbiano, G. \& Kim, D.-W. 1995b, ApJ, 442, 523

Eskridge, P. B., Fabbiano, G. \& Kim, D.-W. 1995c, ApJ, 448, 70 

\reference{} Faltenbacher, A. \& Mathews, W. G. 2005, MNRAS,
358, 139


Gastaldello, F., Buote, D. A., Humphrey, P. J., 
Zappacosta, L., Bullock, J. S., Brighenti, F., 
Mathews, W. G. 2006, ApJ (submitted) (astro-ph/0610134)

\reference{} Humphrey, P. J., Buote, D. A., Gastaldello, F., 
Zappacosta, L, Bullock, J. S., Brighenti, F. \& Mathews, W. G. 
2006, ApJ (in press) (astro-ph/0601301)

\reference{} Humphrey, P. J., Buote, D. A. 
\& Canizares, C. R. 2004, ApJ, 617, 1047

\reference{} Kim, D.-W. \& Fabbiano, G. 2004, ApJ, 611, 846

\reference{} Kravtsov, A. V., Nagai, D., Vikhlinin, A. A. 2005,
ApJ, 625, 588

\reference{} Lin, Y.-T. \& Mohr, J. J. 2004, ApJ, 617, 879

\reference{} Lin, Y.-T., Mohr, J. J. \& 
Stanford, S. A. 2003, ApJ, 591, 749

\reference{} Mathews, W. G., Faltenbacher, A., Brighenti, F. 
\& Buote, D. A. 2005, ApJ, 634, L137

\reference{} Mathews, W. G. \& Brighenti, F. 2003, 
Ann. Rev. Astron. \& Ap. 41, 191

\reference{} Mathews, W. G. \& Brighenti, F. 1998b, ApJ, 503, L15

\reference{} Mathews, W. G. \& Brighenti, F. 1998a, ApJ, 493, L9

\reference{} O'Sullivan, E. \& Ponman, T. J. 2004, MNRAS, 354, 935

\reference{} Reiprich, T. H. \& B\"ohringer, H. 2002, ApJ, 567, 716

Sarazin, C. L., Irwin, J. A. \& Bregman, J. N. 2001, ApJ, 556, 533


\reference{} Spergel, D. N. et al. 2006, ApJ (submitted) (astro-ph/0603449)

\reference{} Sun, M., Jerius, D. \& Jones, C. 2005, ApJ, 633, 165

\reference{} Sun, M., Forman, W., 
Vikhlinin, A., Hornstrup, A., Jones, C. \&
Murray, S. S. 2004, ApJ, 612, 805

\reference{} Vikhlinin, A. et al. 2006, ApJ, 640, 691

\reference{} Yang, X., Mo, H.-J., Kauffmann, G.  \& Chu, Y.-Q. 2003, 
MNRAS, 358, 217

\end{references}
\end{document}